\documentclass[prd,preprint,aps,amsmath,amssymb,showpacs,superscriptaddress,nofootinbib]{revtex4-2}
\usepackage{float}
\usepackage{graphicx} 
\usepackage{dcolumn}
\usepackage{bm}
\usepackage{hyperref}
\usepackage{color}
\usepackage[normalem]{ulem}
\usepackage{comment}
\usepackage{diagbox}
\renewcommand\[{\begin{equation}} 
\renewcommand\]{\end{equation}}

\newcommand{\GeV}{\rm GeV}

\usepackage{ulem}

\begin{document}
\preprint{PITT-PACC-2605}
\preprint{UT-WI-09-2026}
\preprint{MI-HET-881}

\title{Phasing out Dark Matter Isocurvature \\ with Thermal Misalignment}

\author{Brian Batell}
\email{batell@pitt.edu}
\affiliation{Pittsburgh Particle Physics, Astrophysics, and Cosmology Center, \\ 
\mbox{Department of Physics and Astronomy, University of Pittsburgh, Pittsburgh, USA}}
\author{Akshay Ghalsasi}
\email{aghalsasi@fas.harvard.edu}
\affiliation{Pittsburgh Particle Physics, Astrophysics, and Cosmology Center, \\ 
\mbox{Department of Physics and Astronomy, University of Pittsburgh, Pittsburgh, USA}}
\affiliation{Jefferson Physical Laboratory, Harvard University, Cambridge, MA 02138, USA}
\author{Subhajit Ghosh}
\email{sghosh@utexas.edu}
\affiliation{Texas Center for Cosmology and Astroparticle Physics, Weinberg Institute, \\
\mbox{Department of Physics, The University of Texas at Austin, Austin, TX  78712, USA}}

\author{Mudit Rai}
\email{muditrai@tamu.edu}
\affiliation{Pittsburgh Particle Physics, Astrophysics, and Cosmology Center, \\ 
\mbox{Department of Physics and Astronomy, University of Pittsburgh, Pittsburgh, USA}}
\affiliation{Mitchell Institute for Fundamental Physics and Astronomy,\\
\mbox{Department of Physics and Astronomy, Texas A\&M
University, College Station, USA}}

\date{\today}

\begin{abstract}

Thermal misalignment provides an alternative to the standard misalignment mechanism for the cosmological production of scalar dark matter. 
In this framework, feeble couplings to particles in the thermal bath generate a finite-temperature potential that drives the scalar towards large field values early in the radiation era, dynamically inducing the misalignment before the onset of scalar oscillations. 
As a result, the relic abundance is controlled primarily by particle masses and couplings rather than the initial field value. 
As a light spectator field, the scalar acquires inflationary fluctuations that are uncorrelated with the adiabatic curvature mode, generically sourcing isocurvature perturbations.
We show that, unlike standard misalignment, where light scalars are strongly constrained by cosmic microwave background bounds on dark matter isocurvature for high-scale inflation, thermal misalignment can naturally suppress the isocurvature signal. This occurs through a novel late-time phase offset between the  background zero mode and the superhorizon perturbations, which reduces the final dark matter density contrast.
Thermal misalignment therefore provides a new and generic route to isocurvature-safe scalar dark matter.
\end{abstract}

\maketitle

\newpage

%%%%%%%%%%%%%%%%%%%%%%%%%%%%%%%%%%%%%%%%
%%%%%%%%%%%%%%%%%%%%%%%%%%%%%%%%%%%%%%%%
\section{Introduction}

Deciphering the nature of dark matter (DM), which accounts for about a quarter of the energy budget of the Universe~\cite{Planck:2018vyg}, remains one of the central goals of particle physics and cosmology. While the existence of DM is well established through a broad array of astrophysical and cosmological observations, its microscopic identity remains obscure, from its basic dynamical properties to the origin of its cosmic abundance.

An appealing DM candidate is an ultralight scalar field $\phi$ with very weak couplings to other fields. The scalar relic abundance is generically generated by the misalignment mechanism~\cite{Preskill:1982cy,Abbott:1982af,Dine:1982ah}: the field is initially displaced from its minimum, is frozen by Hubble friction in the early radiation era, and begins coherent oscillations when the Hubble rate falls below its mass. The oscillating condensate redshifts as non-relativistic matter and therefore acts as cold DM. In the standard misalignment scenario, the scalar relic abundance is determined by its initial displacement $\phi_i$ and does not depend on its couplings to other particles in the thermal bath. Popular realizations of this scenario include the QCD axion, axion-like particles, as well as dilatons and other light scalar fields (see Ref.~\cite{Antypas:2022asj} for a comprehensive review). 

In this work we study the thermal misalignment mechanism~\cite{Piazza:2010ye,Batell:2021ofv,Batell:2022qvr}, an alternative to the standard misalignment mechanism for the cosmological production of the ultralight scalar DM. In this framework, the scalar couples feebly to states in the thermal bath through couplings that break its shift  symmetry.  The resulting finite-temperature correction to the effective potential in the early radiation era pushes the field toward its high-temperature minimum at large field values, thereby generating the misalignment dynamically. If the initial field value is small compared to the thermally induced misalignment, the late-time oscillation amplitude and resulting  relic abundance becomes insensitive to initial conditions and is instead set by the particle masses and scalar couplings to particles in the bath. Thermal misalignment therefore yields a sharp prediction for the model parameters that reproduce the observed relic density. Thermal misalignment has been further studied in several scenarios with scalar DM coupled to SM particles~\cite{Fardon:2003eh,Brzeminski:2020uhm,Chun:2021uwr,Murgui:2023kig,Plestid:2024kyy,Cyncynates:2024ufu,Cyncynates:2024bxw}, and the general phenomenon has also been explored in other contexts~\cite{Buchmuller:2003is, Buchmuller:2004xr,Moroi:2013tea,Lillard:2018zts,Croon:2022gwq,Cheek:2022yof,Knapp-Perez:2025tns}. 

A key issue that has not yet been studied in the thermal misalignment scenario is the generation of isocurvature perturbations and their associated constraints. In the standard misalignment scenario, ultralight scalar DM behaves as a spectator field during inflation and acquires quantum fluctuations uncorrelated with the adiabatic curvature perturbation generated by the inflaton. 
These fluctuations source DM isocurvature perturbations, which can leave observable imprints on the CMB anisotropies~ \cite{Seckel:1985tj,Linde:1987bx,Lyth:1989pb,Turner:1990uz,Lyth:1992tx,Fox:2004kb,Hertzberg:2008wr,Langlois:2010xc,Langlois:2004nn,Kobayashi:2013nva,Tenkanen:2019aij,Kalia:2025uxg,Allali:2025pja}. 
For a light field in quasi-de Sitter inflation, the fluctuation amplitude is $\delta \phi_i \sim H_I/(2\pi)$.  Since the relic abundance from standard misalignment scales as 
$\rho_\phi \propto m_\phi^2 \phi_i^2$ with $m_\phi$ the scalar mass, inflationary fluctuations induce isocurvature of order
$\delta \rho_\phi/\rho_\phi \sim 2 \delta \phi_i/\phi_i   \sim H_I/ \pi \phi_i$.
CMB observations, including Planck~\cite{Planck:2018jri} , have uncovered no evidence for primordial isocurvature and therefore place an upper bound on the inflationary Hubble scale $H_{I} \leq 1.7\times 10^{7} ~{\rm GeV}  \times (m_{\phi}/{\rm eV})^{-1/4}$.
Reproducing the observed amplitude of primordial curvature perturbations with simple, natural inflationary potentials often points to a relatively high inflation scale, well above the bound discussed above, though still subject to the CMB constraint of $H_{I} \lesssim 4.7 \times 10^{13} \, {\rm GeV}$  from the non-observation of primordial tensor modes~\cite{BICEP:2021xfz}.
Moreover, although low-scale inflation remains viable, many well-studied low-scale constructions are small-field models, which are often less robust to inhomogeneous initial conditions than large-field models~\cite{Clough:2016ymm}. This tension between high-scale inflation and light scalar DM, as implied by CMB isocurvature bounds, has been emphasized repeatedly in the literature, particularly for axions~\cite{Hertzberg:2008wr,Seckel:1985tj,Fox:2004kb,Lyth:1992tx, Lyth:1989pb, Kawasaki:1997ct, Beltran:2006sq, Kawasaki:2007mb,Mack:2009hv}.
Several classes of solutions have been proposed for axion DM, including mechanisms that rely on making the axion heavy during inflation to suppress the amplitude of isocurvature perturbations~\cite{Chakraborty:2025lyp, Dvali:1995ce, Jeong:2013xta, Choi:2015zra, Co:2018phi, Heurtier:2021rko, Dine:2004cq, Higaki:2014ooa, Dine:2014gba, Kawasaki:2015lea, Takahashi:2015waa, Kearney:2016vqw, Buen-Abad:2019uoc, Kawasaki:2015lpf,Nomura:2015xil, Kawasaki:2017xwt, Fischler:1983sc} or that reduce the axion decay constant following inflation~\cite{Linde:1991km,Kasuya:1996ns,Kasuya:1997td, Folkerts:2013tua, Kawasaki:2013iha, Chun:2014xva, Fairbairn:2014zta, Nakayama:2015pba, Harigaya:2015hha, Kobayashi:2016qld, Allali:2022yvx, Graham:2018jyp, Conlon:2022pnx, Co:2019jts} (see \cite{Chakraborty:2025lyp} for more discussion).

Since thermal misalignment alters the post-inflationary evolution of the scalar field, it is natural to ask whether the same isocurvature constraints apply in that scenario.
Our main goal in this work is to study the evolution of isocurvature perturbations in the thermal misalignment scenario and the resulting constraints on parameter space. To this end, we focus on the simple representative model studied in Ref.~\cite{Batell:2021ofv}, consisting of an ultralight scalar DM field coupled to a fermion that is in thermal equilibrium with the SM plasma. We find that, compared to standard misalignment, the isocurvature can be dynamically suppressed in regions of parameter space where the scalar starts oscillating before the thermal potential has switched off. Intuitively, standard misalignment corresponds to a field released from rest, while thermal misalignment can release the field with a sizable initial velocity. This produces an approximately $-\pi/2$ phase shift in the late-time zero mode oscillation, leading to a relative phase offset with the superhorizon perturbation and thus a suppressed DM density contrast. This provides a novel mechanism for suppressing DM isocurvature, distinct from the existing approaches in the axion literature (see previous paragraph), which instead rely on suppressing the amplitude of the primordial isocurvature perturbations.

The remainder of the paper is organized as follows. In Section~\ref{sec:tm}, we introduce the model with a scalar--fermion Yukawa coupling and outline the basic setup for the evolution equations governing the background and superhorizon perturbations in the perturbed expanding universe. In Section~\ref{sec:background}, we develop the solution for the late-time evolution of the scalar zero mode and relate the scalar--fermion coupling to the present-day DM abundance. In Section~\ref{sec:perturbations}, we solve for the late-time behavior of the superhorizon perturbations. In Section~\ref{sec:isocurvature}, we combine these results to determine the modification of large-scale isocurvature in thermal misalignment. We then discuss the implications in Section~\ref{sec:results}, where we show that DM isocurvature can be suppressed.
Finally, we conclude and provide an outlook in Section~\ref{sec:conclusions}.
%%%%%%%%%%%%%%%%%%%%%%%%%%%%%%%%%%%%%%%%
%%%%%%%%%%%%%%%%%%%%%%%%%%%%%%%%%%%%%%%%
\section{Thermal Misalignment: Model and Setup}
\label{sec:tm}

We are interested in the production of DM via the thermal misalignment mechanism and the accompanying generation of isocurvature perturbations. 
To this end, we work in a perturbed Friedmann–Robertson–Walker (FRW) spacetime and adopt the Newtonian gauge. The metric takes the form \cite{Kodama:1984ziu,Mukhanov:1990me,Ma:1995ey,Hu:2003hjx}
\begin{equation}
ds^2 =  -[1+2 \Psi(t, {\bf x})] dt^2 + a^2(t)[1+2 \Phi(t, {\bf x})]\delta_{ij} dx^i dx^j \, ,
\end{equation}
where $a(t)$ is the scale factor and $\Psi$, $\Phi$ are the scalar metric perturbations.

Throughout this work, we focus on a prototypical model consisting of a real scalar field $\phi$, which serves as the DM candidate, coupled feebly to a Dirac fermion $\psi$ via a Yukawa interaction. The Lagrangian is given by
\begin{align}
    \label{eq:phithermal}
    -\mathcal{L} \supset \frac{1}{2} m^{2}_{\phi} \phi^{2} + m_{\psi}\left(1- \frac{\beta \phi}{M_{\rm pl}} \right)\bar{\psi} \psi \, ,
\end{align}
where $m_\phi$ ($m_\psi$) denotes the scalar (fermion) mass, $M_{\rm pl} = 2.4 \times 10^{18}$ GeV is the reduced Planck mass, and the Yukawa coupling is parameterized as $-\beta m_\psi /M_{\rm pl}$ with $\beta$ a real dimensionless parameter.

We assume the fermion $\psi$ is in thermal equilibrium with the Standard Model (SM) plasma during the radiation era. 
The scalar field thus evolves under the influence of an effective potential $V(\phi,T)$, which includes the tree-level mass term in Eq.~(\ref{eq:phithermal}), $V_0(\phi) = \tfrac{1}{2} m_\phi^2 \phi^2$, along with a finite-temperature correction arising from the thermal free energy density of $\psi$, 
\begin{equation}
\label{eq:delta-VT}
 \delta V(\phi,T) =  - \frac{g_{\psi}}{2 \pi^{2}} T^{4} J_{F}\left[\frac{m_\psi^2(\phi)}{T^2}\right] \, ,
\end{equation}
where $g_\psi = 4$ accounts for the fermion spin degrees of freedom, $m_\psi(\phi) = m_\psi (1-\beta \phi/M_{\rm pl})$ is the effective fermion mass in the scalar background, and 
\begin{equation}
J_F(w^2) = \int_0^\infty dx x^2 \log\left[ 1 + e^{-\sqrt{x^2+w^2}}  \right] \, ,
\end{equation}
with $w^2 =(m_\psi^2/T^2)(1-\beta \phi_0/M_{\rm pl})^2$. 

Aside from its coupling to $\psi$ in Eq.~(\ref{eq:phithermal}), we assume $\phi$ has negligible couplings to other particles in the thermal bath. Thus, although $\psi$ is in thermal equilibrium, $\phi$ never thermalizes, and thermal freeze-in produces a negligible scalar relic abundance, owing to the extremely suppressed scalar–fermion coupling.

We decompose the scalar field into a spatially homogeneous, time-dependent background $\phi_0(t)$ and a perturbation $\phi_1(t, {\bf x})$, 
\begin{equation}
\phi(t, {\bf x}) = \phi_0(t)+\phi_1(t, {\bf x}) \, .
\end{equation}
The SM bath temperature is also perturbed according to $T(t, {\bf x}) \rightarrow  T(t)+\delta T(t, {\bf x})$. Expanding the equations of motion to linear order in perturbations yields the evolution equations for $\phi_0$ and $\phi_1$. 
For the perturbations, we track the evolution of the Fourier modes, e.g., $\phi_{1}(t,{\bf k})$ defined via 
\begin{equation}
\label{eq:fourier}
\phi_1(t,{\bf x})  = \int \! \frac{d^3k}{(2\pi)^3} e^{i {\bf k}\cdot {\bf x}} \phi_{1}(t,{\bf k}) \, .
\end{equation}
The equations governing $\phi_0(t)$ and $\phi_{1}(t,{\bf k})$ are (overdots denote derivatives with respect to coordinate time $t$)
\begin{align}
\label{eq:phi0}
& \ddot \phi_0 + 3 H \dot\phi_0 + V_\phi   = 0 \, , \\
& \label{eq:phi1}
\ddot\phi_1 + 3 H \dot \phi_1 + \left( \frac{k^2}{a^2}+V_{\phi\phi}  \right) \phi_1 + (-\dot\Psi + 3\dot \Phi) \dot\phi_0 + 2 \Psi V_{\phi}  +V_{\phi T}  \delta T  = 0 \, ,
\end{align}
where $H = \dot a/a$ is the Hubble parameter,  $V_\phi \equiv \partial V/\partial\phi$, $V_{\phi\phi}  \equiv \partial^2V/\partial\phi^2$, and $V_{\phi T}  \equiv \partial^2V/\partial\phi\partial T$.
The various derivatives appearing in the evolution equations are given by
\begin{align}
{V}_\phi & =  m_\phi^2 \phi_0 + \frac{g_\psi}{\pi^2} \frac{\beta m_\psi^2 T^2 }{M_{\rm pl}} \left(1-\frac{\beta \phi_0}{M_{\rm pl}}\right) J'_F(w^2) \, ,  \\
{V}_{\phi\phi} & =  m_\phi^2 - \frac{g_\psi}{ \pi^2 } \frac{\beta^2 m_\psi^2 T^2}{M_{\rm pl}^2} \left[ J'_F(w^2) +2 w^2 J''_F(w^2)  \right] \, ,\\
 {V}_{\phi T} & =  \frac{2 g_\psi}{ \pi^2 } \frac{\beta m_\psi^2 T }{M_{\rm pl}}  \left(1-\frac{\beta \phi_0}{M_{\rm pl}}\right)  \left[ J'_F(w^2) - w^2 J''_F(w^2)  \right] \, ,
\end{align}

At late times, the scalar field energy density is decomposed into a background component and a perturbation:
\begin{align}
\label{eq:energy0}
\rho_\phi & = \frac{1}{2} \dot \phi^{\,2}_0+  \frac{1}{2} m_\phi^2 \phi_0^2 \, ,  \\
\label{eq:energy1}
\delta\rho_\phi & = \dot\phi_0 \dot\phi_1 -  \Psi  \,\dot \phi^{\,2}_0 + m_\phi^2 \phi_0 \phi_1 \, .
\end{align}
These will be used  in the computation of the DM relic abundance and isocurvature perturbations.

We will be interested in the evolution of superhorizon modes, $k \ll a H$, during the radiation domination era, during which $H = 1/ (2 t) =  \gamma\, T^2/M_{\rm pl}$ with $\gamma(T) = \sqrt{\pi^2 g_{*}(T)/90}$ and $g_{*(S)}(T)$ the effective number of relativistic (entropy) degrees of freedom. 
We therefore neglect the $k^2$ term in Eq.~(\ref{eq:phi1}), take $\Phi \approx -\Psi$ to be constant, $\dot\Psi = \dot \Phi \approx 0$,  and $\delta T/T \approx - \Psi/2$. 
These approximations are valid during radiation domination in the absence of significant anisotropic stress and non-adiabatic pressure perturbation~\footnote{Although, $\phi$ can induce non-adiabatic pressure perturbation~\cite{Malik:2002jb}, we neglect those effects in total curvature perturbation since $\phi$ carry a very small energy density compared to the radiation bath. This is consistent with our treatment of $H = 1/(2t)$ where we have only kept the radiation contribution.}~\cite{Kodama:1984ziu,Mukhanov:1990me,Ma:1995ey,Hu:2003hjx,Malik:2002jb}.
It is also convenient to introduce the following dimensionless variables,
\begin{equation}
\label{eq:dimlesseqs}
\varphi = \frac{\phi}{M_{\rm pl}}\, , ~~~~~~ x = m_\phi t\, , ~~~~~~\kappa = \frac{m_\phi M_{\rm pl}}{m_\psi^2}\, , ~~~~~~ x_\psi = \frac{m_\psi}{H( T = m_\psi)} = \frac{\kappa }{2 \gamma} \, .
\end{equation}
Here, $\varphi$ is the field value relative to $M_{\rm pl}$, $x$ is the dimensionless time variable, while $x_\psi$ corresponds to the ``time'' at which the temperature is equal to the fermion mass, $T = m_\psi$. The latter is proportional to the dimensionless combination of mass parameters $\kappa$. 
The evolution equations for the background and superhorizon perturbations in radiation domination, as well as the corresponding energy density equations, become
\begin{align}
\label{eq:phi0-x}
& \frac{ d^2\varphi_0}{dx^2} + \frac{3}{2x}\frac{d\varphi_0}{dx} +{\cal V}_\varphi= 0 \, , \\
\label{eq:phi1-x}
&\frac{d^2 \varphi_1}{dx^2}  + \frac{3}{2x}  \frac{d\varphi_1}{dx} 
+ {\cal V}_{\varphi\varphi}   \varphi_1  + 2 \Psi {\cal V}_{\varphi}  + \Psi x {\cal V}_{\varphi x} = 0 \, , \\
\label{eq:energy0-x}
& \rho_\phi =m_\phi^2 M_{\rm pl}^2\left[ \frac{1}{2} \left(\frac{ d\varphi_0}{dx}\right)^2 +  \frac{1}{2}  \varphi_0^2 \right] \, ,  \\
\label{eq:energy1-x}
& \delta\rho_\phi = m_\phi^2 M_{\rm pl}^2 \left[\frac{ d\varphi_0}{dx} \, \frac{ d\varphi_1}{dx} - 
\Psi   \left(\frac{ d\varphi_0}{dx}\right)^2 + \varphi_0 \varphi_1\right] \, ,
\end{align}
where we have defined the dimensionless effective potential,
\begin{align}
{\cal V}(\varphi,x) & \equiv \frac{V(\phi = M_{\rm pl}\, \varphi, \,T = m_\psi \sqrt{x_\psi/ x}\,)}{m_\phi^2  M_{\rm pl}^2}  \nonumber \\
& = \frac{1}{2} \varphi^2 - \frac{g_\psi}{2 \pi^2 } \frac{1}{\kappa^2} \frac{x_\psi^2}{x^2} J_F(w^2) \, . 
\end{align}
where $w^2 =(x/x_\psi)(1-\beta \varphi_0)^2$.
Then, the various derivatives appearing in the evolution equations are given by
\begin{align}
\label{eq:V-varphi}
{\cal V}_\varphi & =  \varphi_0 + \frac{g_\psi}{\pi^2 } \frac{\beta}{\kappa^2}  \frac{x_\psi}{x} (1-\beta \varphi_0) J'_F(w^2) \, ,  \\
\label{eq:V-varphi-varphi}
{\cal V}_{\varphi\varphi} & =  1 - \frac{g_\psi}{ \pi^2 }  \frac{\beta^2}{\kappa^2}  \frac{x_\psi}{x}  \left[ J'_F(w^2) +2 w^2 J''_F(w^2)  \right] \, ,\\
\label{eq:V-varphi-x}
{\cal V}_{\varphi x} & =   -\frac{g_\psi}{\pi^2 }\frac{\beta}{\kappa^2}  \frac{x_\psi}{x^2}(1-\beta \varphi_0) \left[ J'_F(w^2) - w^2 J''_F(w^2)  \right] \, .
\end{align}

In order to solve Eqs.~(\ref{eq:phi0-x},\ref{eq:phi1-x}), we must specify the initial conditions at time $x_{i} \simeq  2m_{\phi}/H_{I} \ll 1$ corresponding to the onset of the radiation era, where $H_I$ is the inflationary Hubble scale and we have assumed instantaneous reheating. For the zero mode, assuming a minimal period of inflation with $N \sim 60$ e-folds, the comoving scale corresponding to our observable universe exits the horizon near the onset of inflation, so the background field value is simply set by the pre-inflationary initial condition, $\varphi_{0} (x_{i}) \equiv \varphi_{0i}$, which we treat as a free parameter, while the initial velocity is negligible,  $d\varphi_{0}(x_i)/dx = 0$.~\footnote{For a sufficiently long period of inflation, $N \gg H_I/m_{\varphi}²$, stochastic inflationary dynamics drives the scalar field to an equilibrium distribution with typical field value $\varphi_{0i} \sim H_I^2/m_\phi$~\cite{Starobinsky:1994bd,Graham:2018jyp,Takahashi:2018tdu}. For small $H_I$, this can dynamically generate the small initial field values required for the thermal misalignment mechanism. However, since observationally significant isocurvature perturbations require a sufficiently large $H_I$ we do not focus on this case here.} While in standard misalignment the relic abundance is set by a sizable initial field value $\varphi_{0i}$, the thermal misalignment regime is instead characterized by small initial field values in comparison to the eventual oscillation amplitude that is dynamically set in the early radiation era, so that in practice $\varphi_{0i} \simeq 0$.

The initial conditions for the superhorizon perturbations are set during inflation with the Bunch-Davies vacuum. 
Since $m_\phi \ll H_I$, in the de Sitter limit (neglecting slow-roll corrections) we have $|\varphi_{1}(x_{i})|\equiv |\varphi_{1,i}| = H_{I}/(2 M_{\rm pl} k^{3/2})$ and $d\varphi_{1}(x_i)/dx = 0$, yielding the characteristic nearly scale invariant power spectrum for a light spectator field (see Section~\ref{sec:isocurvature} below for further details).

It is worth remarking that, in addition to the finite-temperature correction in Eq.~(\ref{eq:delta-VT}), the scalar effective potential also receives a zero-temperature one-loop contribution, namely the Coleman--Weinberg potential~\cite{Coleman:1973jx}. Throughout this work, we assume that the full zero-temperature effective potential is nevertheless well approximated by the simple quadratic form in Eq.~(\ref{eq:phithermal}). For small scalar masses $m_\phi$ and and sizable couplings $\beta$, this requires tuning of both the scalar mass term and the quartic coupling, reflecting the usual naturalness problem of light scalars. A rough criterion for radiative stability is $\beta \lesssim 4\pi \kappa$; see Fig.~\ref{fig:TMplot}. In the regions where this condition is violated, fine-tuning of the scalar potential is required. Although we do not pursue this question here, it would be interesting to explore model-building mechanisms that protect such light, weakly coupled scalars; see Refs.~\cite{Hook:2018jle,Brzeminski:2020uhm} for recent progress in this direction.

Our analysis adopts a phenomenological approach in which the scalar condensate evolves in a time-dependent effective potential. While this provides a leading description of thermal misalignment dynamics, it neglects potentially important nonequilibrium effects associated with the full real-time initial-value problem, including derivative terms in the effective action, particle production, and backreaction. Recent work has emphasized that, in Yukawa-coupled systems, these effects can qualitatively modify condensate evolution beyond what is captured by the effective potential alone~\cite{Herring:2025xkc}. While the finite-temperature, expanding background setting relevant for thermal misalignment differs from the zero-temperature Minkowski setup studied there, it would be interesting to investigate analogous nonequilibrium corrections in the present context. Relatedly, our analysis neglects possible thermal damping, a dissipative correction that can arise when the scalar background modulates the masses of particles in the thermal bath~\cite{Yokoyama:1998ju,Banerjee:2025qyg}. As the scalar evolves, these species are driven slightly out of equilibrium, and their relaxation back to equilibrium induces an additional friction-like term beyond the leading thermal effective potential. We have checked that, for the couplings and fermion masses relevant to our study, the associated damping rate is negligible and does not impact the scalar evolution or our results.

In the next two sections, we develop approximate analytic solutions to the evolution equations for the zero mode (\ref{eq:phi0-x}) and the superhorizon perturbations (\ref{eq:phi1-x}).

%%%%%%%%%%%%%%%%%%%%%%%%%%%%%%%%%%%%%%%%
%%%%%%%%%%%%%%%%%%%%%%%%%%%%%%%%%%%%%%%%
\section{Background Evolution and Dark Matter Abundance}
\label{sec:background}

In this section we study the dynamical misalignment of the scalar background $\varphi_0(x)$ driven by the finite temperature effective potential and estimate the resulting DM relic abundance. We also determine the model parameters (coupling and masses) leading to the observed relic density. 
This section thus serves as a review of the thermal misalignment mechanism introduced in Ref.~\cite{Batell:2021ofv}. While our method for solving the zero-mode evolution differs from that used there, we recover the results.

To this end, we solve the evolution equation for $\varphi_0(x)$, Eqs.~(\ref{eq:phi1-x},\ref{eq:V-varphi}) with initial conditions $\varphi_0(x)\vert_{x_i} = \varphi_{0i}$, 
$d\varphi_0/dx \vert_{x_i} = 0$ at the beginning of the radiation era corresponding to $x = x_i \simeq  2m_{\phi}/H_{I}$. In practice, since $m_\phi \ll H_I$, we have $x_i \simeq 0$. 
As mentioned above, the  thermal misalignment regime corresponds to small initial field values in comparison to the eventual oscillation amplitude. Nevertheless, we will retain $\varphi_{0i} \neq 0$ in constructing the zero mode solution for two reasons. First, this will allow us to recover the standard misalignment limit, $\beta \rightarrow 0$ with $\varphi_{0i}\neq 0$, allowing for a clear comparison with the thermal misalignment regime, $\beta \neq 0$ and $\varphi_{0i}\rightarrow 0$. Second, in the next section this more general solution will allow us to directly extract the evolution of the superhorizon perturbations.

As we will see shortly, for the vast majority of the viable parameter space, the correct relic abundance is obtained for couplings $\beta$ and background field values $\varphi_0$ that  satisfy the condition
\begin{equation}
\beta \varphi_0 \ll 1 \, .
\end{equation}
We use this fact to develop a semi-analytic solution to the evolution equation. 
To this end, we expand the derivative of the potential  ${ {\cal V}_\varphi}$ (\ref{eq:V-varphi}) in the background evolution equation~(\ref{eq:phi0-x}) to first order in $\beta \varphi_0$, leading to 
\begin{equation}
\label{eq:phi0-x-expand}
\frac{d^2\varphi_0}{dx^2} + \frac{3}{2x}\frac{d\varphi_0}{dx} + \varphi_0  \simeq - {\delta \widetilde {\cal V}_\varphi}(x) -  {\delta \widetilde {\cal V}_{\varphi\varphi}} \varphi_0 \, , 
\end{equation}
where 
\begin{align}
\label{eq:delta-V-varphi}
\delta \widetilde {\cal V}_\varphi(x) &=  \frac{g_\psi}{\pi^2 } \frac{\beta}{\kappa^2}\frac{x_\psi}{x}  J'_F(x/x_\psi) \, ,  \\
\label{eq:delta-V-varphi-varphi}
\delta \widetilde {\cal V}_{\varphi\varphi} & =  - \frac{g_\psi}{ \pi^2 }  \frac{\beta^2}{\kappa^2}  \frac{x_\psi}{x}  \left[ J'_F(x/x_\psi) +2 (x/x_\psi) J''_F(x/x_\psi)  \right] \, .
\end{align}
The second term on the RHS of Eq.~(\ref{eq:phi0-x-expand}) represents a correction of order $\beta \varphi_0$ which we treat as a perturbation. We thus develop a perturbative solution for the background, 
\begin{equation}
\label{eq:background-perturbative}
\varphi_0(x)  \simeq \varphi_0^{(0)}(x) +  \varphi_0^{(1)}(x) \, .
\end{equation}
We emphasize that $\varphi_0^{(1)}$ is \textbf{\em{not}} the inhomogeneous linear perturbation (i.e., $\varphi_{1}$), but is instead a first order correction to the homogeneous background in the $\beta \varphi_{0} \ll 1$ limit.
Plugging Eq.~(\ref{eq:background-perturbative}) into (\ref{eq:phi0-x-expand}),  we find that $\varphi_0^{(0)}$ and $\varphi_0^{(1)}$ evolve according to 
\begin{align}
\label{eq:phi0-x-0}
\frac{d^2\varphi_0^{(0)}}{dx^2} + \frac{3}{2x}\frac{d\varphi_0^{(0)}}{dx} + \varphi_0^{(0)}  & = - {\delta \widetilde {\cal V}_\varphi} ,  \\
\label{eq:phi0-x-1}
\frac{d^2\varphi_0^{(1)}}{dx^2} + \frac{3}{2x}\frac{d\varphi_0^{(1)}}{dx} + \varphi_0^{(1)}  & =  -  {\delta \widetilde {\cal V}_{\varphi\varphi}} \varphi^{(0)}_0, 
\end{align} 
This solution method corresponds to the Born approximation familiar from quantum mechanics. The terms on the right-hand side act as field-independent sources, while the homogeneous equations are readily solved in terms of Bessel functions. We therefore adopt a Green’s function approach to solve Eqs.~(\ref{eq:phi0-x-0}, \ref{eq:phi0-x-1}).

The solution to Eq.~(\ref{eq:phi0-x-0}), subject  to the initial conditions at $x_i \simeq 0$ given by $\varphi^{(0)}_{0}(0) = \varphi_{0i}$, $d\varphi^{(0)}_{0}(0)/dx = 0$, is given by 
\begin{equation}
\label{eq:varphi-0-0}
\varphi^{(0)}_{0}(x) = \varphi_{0i} \, 2^{1/4} \, \Gamma(5/4) \, \frac{J_{1/4}(x)}{x^{1/4}} -  \int_{0}^{x} dy \, {\cal G}(x,y) \, \delta \widetilde {\cal V}_\varphi(y).
\end{equation}
Here $\Gamma(z)$ is the Euler gamma function, while ${\cal G}(x,y)$ is the retarded Green's function, defined by ${\cal G}(x,y) = 0$  for $x < y$ and, for $x > y$, 
\begin{equation}
\label{eq:Greens}
{\cal G}(x,y) = \frac{\pi}{2} \frac{y^{5/4}}{x^{1/4}} \left[ Y_{1/4}(x) J_{1/4}(y)  -J_{1/4}(x) Y_{1/4}(y)  \right],
\end{equation}
with $J_{1/4}(x)$, $Y_{1/4}(x)$  the Bessel functions of order 1/4. The first (second) term in Eq.~(\ref{eq:varphi-0-0}) represents the homogeneous (particular) solution. 

Using Eq.~(\ref{eq:varphi-0-0}) for $\varphi^{(0)}_{0}(x)$, we immediately obtain the solution to Eq.~(\ref{eq:phi0-x-1}),
\begin{equation}
\label{eq:varphi-0-1}
\varphi^{(1)}_{0}(x) = -  \int_{0}^{x} dy \, {\cal G}(x,y) \, \delta \widetilde {\cal V}_{\varphi\varphi}(y) \varphi_0^{(0)}(y).
\end{equation}

The full solution to the background field equation, valid for small $\beta \varphi_0$, is given by the the sum of (\ref{eq:varphi-0-0}) and (\ref{eq:varphi-0-1}) and may be written as 
\begin{align}
\label{eq:varphi-0-sol}
\varphi_0(x)
&= {\cal A}(x,x_\psi)\,\frac{J_{1/4}(x)}{x^{1/4}}
 + {\cal B}(x,x_\psi)\,\frac{Y_{1/4}(x)}{x^{1/4}},
\end{align}
where the coefficients ${\cal A}$ and ${\cal B}$  are given by
\begin{align}
{\cal A}(x,x_\psi)
&= \varphi_{0i}\,2^{1/4}\Gamma(5/4)
+\frac{g_\psi}{4\gamma\pi}\frac{\beta}{\kappa}\,{\cal F}_Y(x,x_\psi)
-\varphi_{0i}\,2^{1/4}\Gamma(5/4)\frac{g_\psi}{2\pi}\frac{\beta^2}{\kappa^2}\,{\cal F}_{JY}(x,x_\psi)  \nonumber \\ &
+\frac{g_\psi^2}{8\gamma\pi^2}\frac{\beta^3}{\kappa^3}\Big[{\cal F}_{YYJ}(x,x_\psi)-{\cal F}_{JYY}(x,x_\psi)\Big],
\label{eq:A-coef} \\
{\cal B}(x,x_\psi)
&= -\frac{g_\psi}{4\gamma\pi}\frac{\beta}{\kappa}\,{\cal F}_J(x,x_\psi)
+\varphi_{0i}\,2^{1/4}\Gamma(5/4)\frac{g_\psi}{2\pi}\frac{\beta^2}{\kappa^2}\,{\cal F}_{JJ}(x,x_\psi) \nonumber \\
&-\frac{g_\psi^2}{8\gamma\pi^2}\frac{\beta^3}{\kappa^3}\Big[{\cal F}_{JYJ}(x,x_\psi)-{\cal F}_{JJY}(x,x_\psi)\Big]. 
\label{eq:B-coef}
\end{align}
The coefficients depend on the integrals ${\cal F}_{J}$, ${\cal F}_{Y}$, ${\cal F}_{JJ}$, ${\cal F}_{JY}$, ${\cal F}_{JYJ}$, ${\cal F}_{JJY}$, ${\cal F}_{YYJ}$, ${\cal F}_{JYY}$, which are defined as
\begin{align}
\label{eq:FJ}
{\cal F}_J(x,x_\psi) & = \int_0^x \!\! dy \, y^{1/4} \, J_{1/4}(y) \, J'_F(y/x_\psi), \\
\label{eq:FY}
{\cal F}_Y(x,x_\psi)  & = \int_0^x \!\! dy \, y^{1/4} \, Y_{1/4}(y) \, J'_F(y/x_\psi), \\
\label{eq:FJJ}
{\cal F}_{JJ}(x,x_\psi) & = x_\psi \int_0^x \!\! dy \, [ J_F'(y/x_\psi) +2 (y/x_\psi) J_F''(y/x_\psi) ] \, J^2_{1/4}(y) , \\
\label{eq:FJY}
{\cal F}_{JY}(x,x_\psi) & = x_\psi \int_0^x \!\! dy \,  [ J_F'(y/x_\psi) +2 (y/x_\psi) J_F''(y/x_\psi) ] \, J_{1/4}(y) \,  Y_{1/4}(y) , \\
\label{eq:FJYJ}
{\cal F}_{JYJ}(x, x_\psi) & = x_\psi \int_0^x \!\! dy \, [ J_F'(y/x_\psi) +2 (y/x_\psi) J_F''(y/x_\psi) ]\, J_{1/4}(y) \,  Y_{1/4}(y)  {\cal F}_J(y,x_\psi) , \\
\label{eq:FJJY}
{\cal F}_{JJY}(x,x_\psi) & = x_\psi \int_0^x \!\! dy \, [ J_F'(y/x_\psi) +2 (y/x_\psi) J_F''(y/x_\psi) ]\, J^2_{1/4}(y) {\cal F}_Y(y,x_\psi) ,  \\
\label{eq:FYYJ}
{\cal F}_{YYJ}(x,x_\psi) & = x_\psi \int_0^x \!\! dy \, [ J_F'(y/x_\psi) +2 (y/x_\psi) J_F''(y/x_\psi) ]\,   Y^2_{1/4}(y)  {\cal F}_J(y,x_\psi) , \\
\label{eq:FJYY}
{\cal F}_{JYY}(x,x_\psi) & = x_\psi \int_0^x \!\! dy \, [ J_F'(y/x_\psi) +2 (y/x_\psi) J_F''(y/x_\psi) ] \,  J_{1/4}(y) \, Y_{1/4}(y) {\cal F}_Y(y,x_\psi) .
\end{align}
For simplicity we have neglected the variation of $g_*$ with temperature in (\ref{eq:varphi-0-sol}). This allows us to obtain semi-analytic solutions in the asymptotic large and small $\kappa$ limits, while only having a small numerical impact on our results.

\begin{figure}
    \centering
    \includegraphics[width=0.49\linewidth]{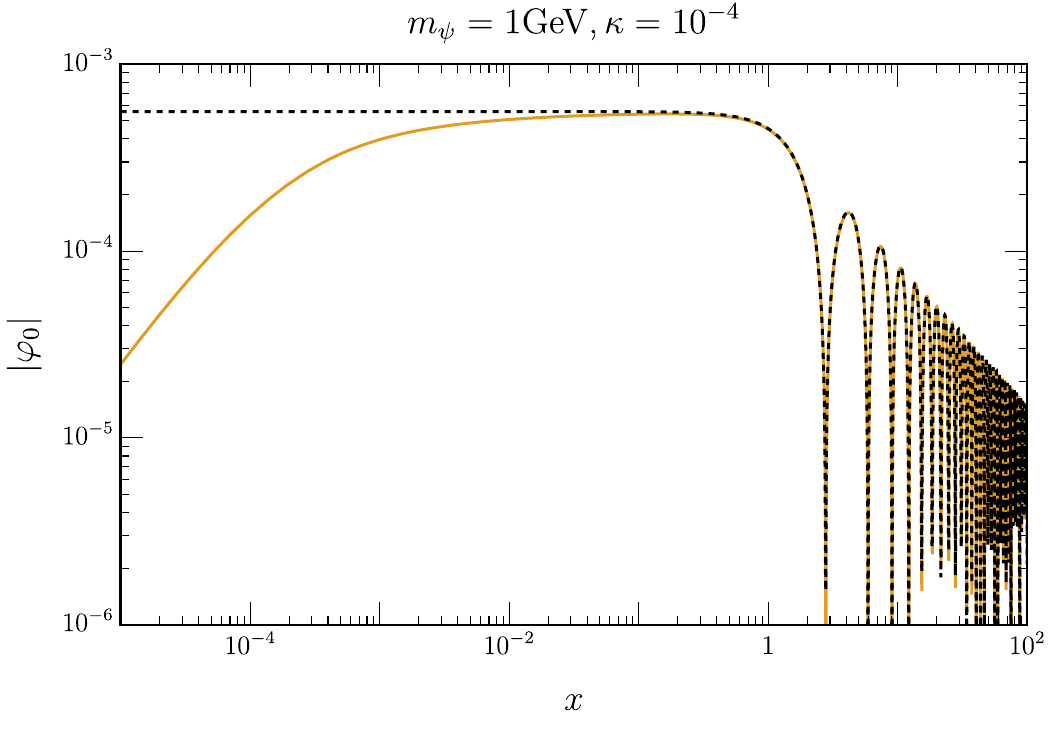}
    \includegraphics[width=0.49\linewidth]{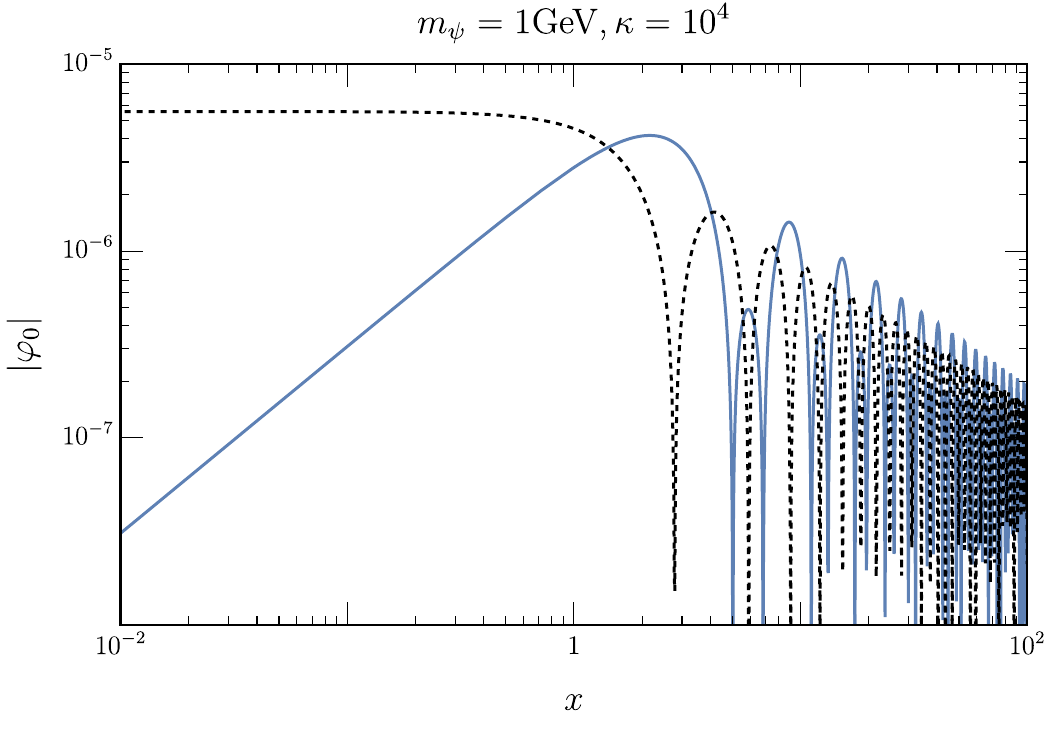}
    \caption{Evolution of $|\varphi_0|$ in the thermal misalignment scenario with $\varphi_{0i} = 0$. The left (right) plot shows the evolution of the scalar in the small (large) $\kappa$ limit where we have chosen the coupling $\beta$ to reproduce the observed relic density (colored lines); see Eqs.~(\ref{eq-phi-osc-tm},\ref{eq:phi-i-DM}). For comparison, we also show the evolution of scalar background in the standard misalignment scenario (black dashed lines). In the small $\kappa$ limit the scalar gets misaligned at small $x$ in the presence of the thermal potential which becomes subdominant at $x \simeq x_{\psi}\ll 1$. The zero mode asymptotes to a constant value and then starts to oscillate, such that at late times the oscillations are in phase with the corresponding evolution in the standard misalignment scenario ($\theta_{0} \simeq 0)$. At large $\kappa$, the thermal potential is dominant around the onset of oscillations and the background field is moving with a large velocity. The late time oscillations of $\varphi_0$ for large $\kappa$ are out of phase with the corresponding evolution in standard misalignment with $\theta_{0}\simeq -\pi/2$.}
    \label{fig:phievol}
\end{figure}

Using the large $x$ expansions for the Bessel functions in Eq.~(\ref{eq:varphi-0-sol}) and taking the upper integration limit $x\rightarrow \infty$ in the integrals  (\ref{eq:FJ}-\ref{eq:FJYY}), the unique late-time solution for the background satisfying the initial conditions can be written as 
\begin{equation}
\label{eq:phi0-late}
\varphi_0(x) \simeq \varphi_{\rm osc} \left(\frac{x_{\rm osc}}{x}\right)^{3/4} \sin\left( x + \frac{\pi}{8} + \theta_0 \right).
\end{equation}
Here $x_{\rm osc} = 3/2$ is the time corresponding to the onset of oscillations, defined by the condition $3 H(x_{\rm osc}) = m_\phi$. Furthermore, $\varphi_{\rm osc}$ and $\theta_0$ are the effective oscillation amplitude and induced phase angle, respectively, given by 
\begin{equation}
\label{eq:varphi-0-osc-amp}
\varphi_{\rm osc} = x_{\rm osc}^{-3/4} (2/\pi)^{1/2}\sqrt{ [A(x_\psi)]^2 + [B(x_\psi)]^2} ,\qquad~~~ \tan \theta_0 = - \frac{ B(x_\psi)}{A(x_\psi)},
\end{equation} 
where $ A(x_{\psi}) \equiv \lim\limits_{x\to \infty}   {\cal A}(x,x_\psi)$,  $ B(x_{\psi}) \equiv \lim\limits_{x\to \infty} {\cal B}(x,x_\psi)$ are the coefficients (\ref{eq:A-coef},\ref{eq:B-coef}) in the limit $x\rightarrow \infty$. The functions $A(x_\psi)$ and $B(x_\psi)$ depend on the integrals in Eqs.~(\ref{eq:FJ}--\ref{eq:FJYY}) evaluated in the $x\rightarrow \infty$ limit. Thus, we define $F_J(x_{\psi}) \equiv \lim\limits_{x\to \infty}   {\cal F}_J(x,x_\psi)$, $F_Y(x_{\psi}) \equiv \lim\limits_{x\to \infty}   {\cal F}_Y(x,x_\psi)$, etc., analogously to $A(x_\psi)$ and $B(x_\psi)$. For general values of $x_\psi$ these integrals must be evaluated numerically. However, one may also obtain asymptotic expressions for these integrals in the small $x_\psi$ and large $x_\psi$ regimes, which are presented in Table~\ref{tab:integrals}. The numerical evolution of the background field $\varphi_0$ is shown in Fig.~\ref{fig:phievol} in the small $\kappa$ and large $\kappa$ limits for thermal misalignment with $\varphi_{0i} \rightarrow 0$, along with a comparison to the standard misalignment case, $\beta \rightarrow 0$, $\varphi_{0i}\neq 0$.
\begin{table}[t]
\centering
\begin{tabular}{| l |  c | c|}
\hline\hline
 & $x_\psi \ll 1$ & $x_\psi \gg 1$ \\
\hline
~$F_J(x_\psi)$~   & ~$-4.25\, x_\psi^{3/2}$~ & ~$-0.338$~ \\
~$F_Y(x_\psi)$~   & ~$2.60\, x_\psi$~        & ~$(0.0791 + 0.0429\,\log x_\psi)/x_\psi$~ \\[4pt]
~$F_{JJ}(x_\psi)$~ & ~$7.89\, x_\psi^{5/2}$~  & ~$-(0.0337 + 0.131\,\log x_\psi)\,x_\psi$~ \\
~$F_{JY}(x_\psi)$~ & ~$-2.58\, x_\psi^{2}$~   & ~$0.206\, x_\psi$~ \\[4pt]
~$F_{JYJ}(x_\psi)$~ & ~$7.57\, x_\psi^{7/2}$~  & ~$-(0.00665 + 0.00470\,\log x_\psi)$~ \\
~$F_{JJY}(x_\psi)$~ & ~$18.1\, x_\psi^{7/2}$~  & ~$-0.0695\, x_\psi$~ \\
~$F_{YYJ}(x_\psi)$~ & ~$-2.57\, x_\psi^{3}$~   & ~$(-0.0420 + 0.0443\,\log x_\psi)\,x_\psi$~ \\
~$F_{JYY}(x_\psi)$~ & ~$-5.70\, x_\psi^{3}$~   & ~$0.0200\, x_\psi$~ \\
\hline\hline
\end{tabular}
\caption{Asymptotic forms of the integrals in the small and large $x_\psi$ regimes.}
\label{tab:integrals}
\end{table}

The DM energy density today is given by
$\rho_{\phi,0} = \tfrac{1}{2} m_\phi^2 M_{\rm pl}^2 \varphi^2_{\rm osc} (T_0/T_{\rm osc})^3 (g_{*S,0}/g_{*S, {\rm osc}})$, where
$T_0 = 2.7$ K is the CMB temperature today and $g_{*S,0} = 3.91$. 
The present-day DM density parameter is then given by $\Omega_\phi = \rho_{\phi,0}/\rho_{c,0}$, where $\rho_{c,0} = 3M_{\rm pl}^2 H_0^2$ the critical density. This should be compared to the observed DM density $\Omega_{\rm DM} = 0.26$ \cite{Planck:2018vyg}.

Before discussing the case of thermal misalignment, we can recover the familiar results for the case of standard misalignment by taking the limit $\beta \rightarrow 0$, in which case Eq.~(\ref{eq:varphi-0-osc-amp}) gives the oscillation amplitude and phase, respectively, of $\varphi_{\rm osc}  \simeq 0.63 \, \varphi_{0i}$ and $\theta_0 = 0$. The DM relic density is then given by 
\begin{align}
\label{eq:relic-standard}
\Omega_\phi & \simeq 0.2  \, \frac{g_{*S,0}}{g^{1/4}_{*,{\rm osc}}}\, \varphi_{0i}^2\, \frac{ \sqrt{m_\phi M_{\rm pl}} \, T_0^3 }{\rho_{c,0} } \\
& \simeq 0.2 \left( \frac{\varphi_{0i}}{4\times 10^{-5}} \right)^2  \left( \frac{m_\phi}{10^{-9}\, {\rm eV}} \right)^{1/2}  \left( \frac{106.75}{ g_{*,{\rm osc}}} \right)^{1/4}. \nonumber
\end{align}
In the case of standard misalignment, we see from Eq.~(\ref{eq:relic-standard}) that the DM abundance is determined by the initial field value. 

\begin{figure}
    \centering
    \includegraphics[width=0.9\linewidth]{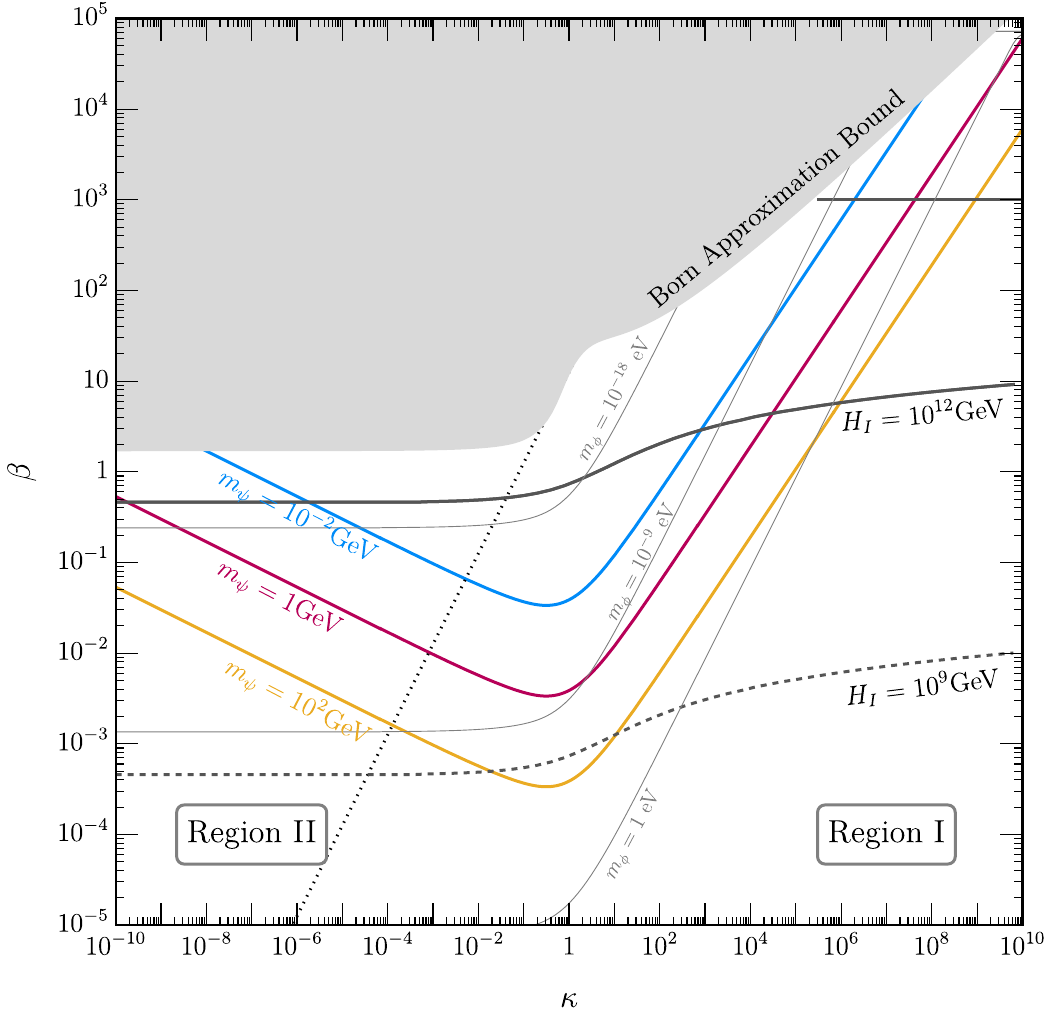}
    \caption{{\it Thermal misalignment parameter space:} Solid colored contours show the values of $\beta$ as a function of $\kappa$ that reproduce the observed relic density for fermion masses $m_{\psi} = 10^{-2} ~\GeV$ (teal), $m_{\psi} =  1~\GeV$ (magenta), $m_{\psi} = 10^{2} ~\GeV$ (yellow).  The shaded region denotes the parameter space in which the Born approximation is no longer valid (see Eq.~(\ref{eq:Born-validity-f})). 
    The dotted black line indicates the approximate naturalness condition $\beta = 4 \pi \kappa$, above which the radiative corrections to the zero temperature scalar potential dominate over its bare value. Thin gray lines represent isocontours of $m_{\phi}$. The region within the solid black lines (above the dashed black line) indicate parameters consistent with the CMB isocurvature constraints for $H_{I} = 10^{12} ~\GeV$ ($H_{I} = 10^{9} ~\GeV$). As we will show in the following sections, thermal misalignment opens up a region of $m_\phi-H_I$ parameter space for large $H_{I}$ that is ruled out for standard misalignment from isocurvature bounds (see Fig. \ref{fig:HIplot}). } 
    \label{fig:TMplot}
\end{figure}

Turning now to the thermal misalignment regime, we take the limit $\varphi_{0i} \rightarrow 0$ so that the misalignment is primarily dynamically generated early in the radiation era. 
For the purposes of determining the relic abundance, it is enough to retain zeroth order solution 
$\varphi_0^{(0)}$, such that
\begin{equation}
\label{eq-phi-osc-tm}
    \varphi_{\rm osc} \simeq \frac{g_\psi}{(6 \pi^2)^{3/4} \gamma} \frac{\beta}{\kappa} \sqrt{[{F}_J(x_\psi)]^2+ [{F}_Y(x_\psi)]^2} \, . 
\end{equation}
As discussed in Ref.~\cite{Batell:2021ofv}, and as can be seen from Fig.~\ref{fig:TMplot}, there are two main regions. 
Region 1 corresponds to $x_\psi \gg 1$, for which ${F}_J(x_\psi) \simeq -0.338$ dominates over ${F}_Y(x_\psi)$. The DM relic density in this region is

\begin{align}
\label{eq:relic-thermal-R1}
\Omega_\phi & \simeq 0.018  \, \frac{g_{*S,0}}{g^{5/4}_{*,{\rm osc}}}\, \frac{\beta^2 m_\psi^4 T_0^3 }{(m_\phi M_{\rm pl})^{3/2} \rho_{c,0}}  \\
& \simeq 0.2 \left( \frac{\beta}{0.02} \right)^2 
 \left( \frac{m_\psi}{1\, {\rm TeV}} \right)^{4}  \left( \frac{1\, {\rm eV}}{ m_\phi} \right)^{3/2} \left( \frac{106.75}{ g_{*,{\rm osc}}} \right)^{5/4} \nonumber,\\
& \simeq 0.2 
\left( \frac{\beta}{0.01} \right)^2  \left( \frac{10^3}{\kappa} \right)^{3/2}
 \left( \frac{m_\psi}{1\, {\rm TeV}} \right)  \left( \frac{106.75}{ g_{*,{\rm osc}}} \right)^{5/4}. \nonumber
\end{align}
where in the first line we have written the relic density as a function of the scalar mass $m_{\phi}$ and in the second line we have replaced $m_{\phi}$ for the dimensionless mass parameter $\kappa$.
Region II corresponds to $x_\psi \ll 1$, for which ${F}_Y(x_\psi) \simeq 2.6 x_\psi$ dominates over ${F}_J(x_\psi)$. The DM relic density in this region is

\begin{align}
\label{eq:relic-thermal-R2}
\Omega_\phi & \simeq 2.4  \, \frac{g_{*S,0}}{g^{9/4}_{*,{\rm osc}}}\, \frac{\beta^2  (m_\phi M_{\rm pl})^{1/2}  T_0^3 }{ \rho_{c,0}}  \\
& \simeq 0.2 \left( \frac{\beta}{4\times 10^{-4}} \right)^2 
 \left( \frac{m_\phi}{10^{-7}\, {\rm eV}} \right)^{1/2}   \left( \frac{106.75}{ g_{*,{\rm osc}}} \right)^{9/4}, \nonumber \\
& \simeq 0.2 
\left( \frac{\beta}{5 \times 10^{-4}} \right)^2  \left( \frac{\kappa }{10^{-4}} \right)^{1/2}
 \left( \frac{m_\psi}{1\, {\rm TeV}} \right)  \left( \frac{106.75}{ g_{*,{\rm osc}}} \right)^{9/4}. \nonumber
\end{align}

It is useful to define the quantity $\varphi_{{\rm DM}}$ corresponding to the effective initial field value prior to the onset of oscillations which yields the observed DM abundance. It is given by
\begin{align}
\label{eq:phi-i-DM}
\varphi_{{\rm DM}} & = \varphi_{\rm osc,{\rm DM}} \frac{3^{3/4} \,\sqrt{\pi} }{2^{3/2}\Gamma(5/4)}  \simeq 
2.2 \, \frac{g^{1/8}_{*,{\rm osc}}}{g_{*S,0}^{1/2}}\left( \frac{\rho_{c,0} \Omega_{\rm DM}}{T^3_{0} \sqrt{m_\phi M_{\rm pl}}} \right)^{1/2} \\
& = 2.5 \times 10^{-7}   \left(\frac{g_{*,{\rm osc}}}{106.75}\right)^{1/8} \left( \frac{1 \,{\rm eV}}{m_\phi}\right)^{1/4} , \nonumber\\
& = 5.5 \times 10^{-5}   \left(\frac{g_{*,{\rm osc}}}{106.75}\right)^{1/8} \left( \frac{1 \,{\rm GeV}}{m_\psi}\right)^{1/2} \kappa^{-1/4}. \nonumber
\end{align}
Using Eqs.~(\ref{eq:phi-i-DM}, \ref{eq-phi-osc-tm}), we can determine the coupling $\beta$ required to generate the observed DM relic density as a function of $\kappa$ for a given fermion mass, as shown in  Fig. \ref{fig:TMplot}.

%%%%%%%%%%%%%%%%%%%%%%%%%%%%%%%%%%%%%%%%
%%%%%%%%%%%%%%%%%%%%%%%%%%%%%%%%%%%%%%%%
\section{Evolution of the perturbations}
\label{sec:perturbations}
Having examined the background evolution and the DM relic abundance, we now turn to the evolution of the superhorizon scalar perturbations. Only the superhorizon evolution is relevant for our purposes, because the CMB-scale modes that constrain isocurvature remain outside the horizon during thermal misalignment. The evolution equation for the superhorizon mode $\varphi_1(x,{\bf k})$ (recall that $x = m_{\phi} t$ is the dimensionless time variable) is given in Eq.~(\ref{eq:phi1-x}) together with Eqs.~(\ref{eq:V-varphi},\ref{eq:V-varphi-varphi},\ref{eq:V-varphi-x}). 
In general, the solution for the superhorizon perturbations can be written as  \cite{Weinberg:2003sw,Allali:2025pja,Kobayashi:2013nva} \footnote{For brevity, we use the same notation for the Fourier-space perturbation and its mode function; the precise meaning should be apparent from the context.} 
\begin{equation}
\label{eq:ansatz}
\varphi_1 = f + \Psi x \frac{d\varphi_0}{dx}.
\end{equation}
Here, the second term represents the adiabatic mode, while the first term, $f$, represents the isocurvature mode.

Plugging Eq.~(\ref{eq:ansatz}) into Eq.~(\ref{eq:phi1-x}), we obtain the following evolution equation for $f$: 
\begin{align}
\label{eq:f-x}
\frac{ d^2f}{dx^2} + \frac{3}{2x}\frac{df}{dx} +{\cal V}_{\varphi\varphi}f= 0 ,
\end{align}
where ${\cal V}_{\varphi\varphi}$ is defined in Eq.~(\ref{eq:V-varphi-varphi}).  For initial conditions at $x_i \simeq 0$ given by $f(0) = \varphi_{1}(0) = \varphi_{1i}$, 
$df(0)/dx = 0$, the solution for the isocurvature mode is given by \cite{Allali:2025pja}
\begin{align}
\label{eq:f-sol}
f(x) & = \varphi_{1i} \frac{\partial \varphi_0(x, \varphi_{0i})}{\partial \varphi_{0i}}  \\
& \simeq  \varphi_{1i} 2^{1/4}\Gamma(5/4) \left[ {\cal C}(x,x_\psi)\,\frac{J_{1/4}(x)}{x^{1/4}} + {\cal D}(x,x_\psi)\,\frac{Y_{1/4}(x)}{x^{1/4}} \right], \nonumber
\end{align}
where we used the solution for the background in Eq.~(\ref{eq:varphi-0-sol}). 
One may verify by direct substitution that Eq.~(\ref{eq:f-sol}) satisfies Eq.~(\ref{eq:f-x}).
The coefficients ${\cal C}$ and ${\cal D}$  are given by 
\begin{align}
{\cal C}(x,x_\psi)
&= 1
- \frac{g_\psi}{2\pi}\frac{\beta^2}{\kappa^2}\,{\cal F}_{JY}(x,x_\psi) , \label{eq:C-coef} \\
{\cal D}(x,x_\psi)
&= \frac{g_\psi}{2\pi}\frac{\beta^2}{\kappa^2}\,{\cal F}_{JJ}(x,x_\psi) ,\label{eq:D-coef}
\end{align}
with integrals ${\cal F}_{JJ}$, ${\cal F}_{JY}$ given in Eqs.~(\ref{eq:FJJ},\ref{eq:FJY})

Using the large $x$ expansions for the Bessel functions in Eq.~(\ref{eq:f-sol}) the late-time solution for $f(x)$ can be written as 
\begin{equation}
\label{eq:f-late}
f(x) \simeq f_{\rm osc} \left(\frac{x_{\rm osc}}{x}\right)^{3/4} \sin\left( x + \frac{\pi}{8} + \theta_1 \right).
\end{equation}
Here $x_{\rm osc} = 3/2$ is again the time corresponding to the onset of oscillations, while $f_{\rm osc}$ and $\theta_1$ are the effective oscillation amplitude and induced phase angle, respectively, given by 
\begin{equation}
\label{eq:f-osc}
f_{\rm osc} = x_{\rm osc}^{-3/4} \varphi_{1i} \frac{2^{3/4} \Gamma(5/4)}{\sqrt{\pi}}  \sqrt{  [C(x_\psi)]^2 +  [D(x_\psi)]^2 },  
\qquad~~ \tan \theta_1 = - \frac{D(x_\psi)}{C(x_\psi)},
\end{equation} 
where $ C(x) \equiv \lim\limits_{x\to \infty}   {\cal C}(x,x_\psi)$,  $ D(x) \equiv \lim\limits_{x\to \infty} {\cal D}(x,x_\psi)$ are the coefficients (\ref{eq:C-coef},\ref{eq:D-coef}) in the limit $x\rightarrow \infty$.

We note that for the Born approximation to hold, the correction to the amplitude of $f$ from the thermal effect should be small, i.e.,
\begin{align}
    \label{eq:Born-validity-f}
    \bigg|\frac{g_\psi}{2\pi}\frac{\beta^2}{\kappa^2}\,{F}_{JJ}(x_\psi)\bigg|,~\bigg|\frac{g_\psi}{2\pi}\frac{\beta^2}{\kappa^2}\,{F}_{JY}(x_\psi)\bigg| \ll 1.
\end{align}
This condition delineates the region of validity of our analysis and is shown in Fig.~\ref{fig:TMplot} \footnote{A similar condition applies to corrections to the zero mode, and we have checked that it gives a numerically similar constraint.}.

%%%%%%%%%%%%%%%%%%%%%%%%%%%%%%%%%%%%%%%%
%%%%%%%%%%%%%%%%%%%%%%%%%%%%%%%%%%%%%%%%
\section{Isocurvature perturbations}
\label{sec:isocurvature}

With the solutions to the scalar DM background and superhorizon perturbations in hand, we can compute the superhorizon isocurvature power spectrum at late times in the radiation domination era. Precision CMB measurements from Planck and Atacama Cosmology Telescope have placed stringent contstraints on the amount of DM isocurvature, which would leave distinctive imprints on the CMB power spectrum~\cite{Planck:2018jri,AtacamaCosmologyTelescope:2025nti}. By computing the effects on the superhorizon evolution for large-scale CMB modes, we can translate these isocurvature constraints to bounds on thermal misalignment parameters. 

The gauge invariant isocurvature perturbation ${\cal S}$ for the scalar field $\phi$~\cite{Bardeen:1983qw,Wands:2000dp,Malik:2002jb},  is 
\begin{equation}
\label{eq:S-gauge-inv}
{\cal S} = 3 (\zeta_\gamma - \zeta_\phi),
\end{equation}
where $\zeta_\alpha$ is the curvature perturbation on uniform $\alpha$-fluid density hypersurfaces~\cite{Wands:2000dp} defined as 
\begin{equation}
\label{eq:zeta-alpha}
\zeta_\alpha = \Phi - H \frac{\delta \rho_\alpha}{d\rho_{\alpha}/dt}.
\end{equation}
Plugging Eq.~(\ref{eq:zeta-alpha}) into (\ref{eq:S-gauge-inv}), we obtain
\begin{align}
{\cal S}  & = 3H \left(\frac{\delta \rho_\phi}{d\rho_{\phi}/dt}- \frac{\delta \rho_\gamma}{d\rho_{\gamma}/dt}\right) =  \frac{3}{2x} \left(\frac{\delta \rho_\phi}{d\rho_{\phi}/dx}- \Psi x\right).
\end{align}
In the second step above we have used the radiation continuity equation $d\rho_\gamma/dt = -4 H \rho_\gamma$, 
the relation $\delta \rho_\gamma/ \rho_\gamma = - 2 \Psi$ for superhorizon modes \cite{Ma:1995ey},  and changed time variables to $x = m_\phi t$.

Furthermore, using Eqs.~(\ref{eq:energy0-x},\ref{eq:energy1-x}) for the background and perturbed $\phi$ energy densities, the $\phi$ continuity equation $d\rho_\phi/dx = - 3/(2x) \rho_\phi$, and the solution for the superhorizon perturbation (\ref{eq:ansatz}), we arrive at the expression 
\begin{align}
\label{eq:S-2}
{\cal S}  = - 2 \frac{ (d\varphi_0/dx)(df/dx) + \varphi_0 \, f  }{  (d\varphi_0/dx)^2 + \varphi_0^2}.
\end{align}
Finally, plugging in the late time solutions for the background and superhorizon perturbations, Eqs.~(\ref{eq:phi0-late}) and (\ref{eq:f-late}), respectively, we obtain the compact expression
\begin{align}
{\cal S} = - 2 \frac{f_{\rm osc}}{\varphi_{\rm osc}} \cos(\theta_0 - \theta_1).
\end{align}
where the oscillation amplitudes and phase angles for the background and perturbation are given in Eqs.~(\ref{eq:varphi-0-osc-amp},\ref{eq:f-late}). 
We now fix the parameters by requiring agreement with the observed DM relic abundance, implying  $\varphi_{\rm osc} \simeq 0.63 \varphi_{{\rm DM}}$, where $\varphi_{{\rm DM}}$ is given  in Eq.~(\ref{eq:phi-i-DM}). Furthermore, we restrict our focus to the regime in which the Born approximation is valid such that $f_{\rm osc} \simeq 0.63 \varphi_{1i}$ (see Eqs.~(\ref{eq:f-osc},\ref{eq:Born-validity-f})) leading to the simple result 
\begin{align}
{\cal S} \simeq - 2 \frac{\varphi_{1i}}{\varphi_{{\rm DM}}} \cos(\theta_0 - \theta_1).
\end{align}

Assuming a standard slow-roll inflationary era which preceded radiation domination, characterized by Hubble parameter $H_I$ and slow-roll parameter $\epsilon \equiv - \dot H_I/H_I^2  \ll 1$, and taking $m_\phi \ll H_I$, perturbations of the scalar field generated during inflation acquire a nearly scale-invariant power spectrum,
\begin{align}
\label{eq:PS-perturbation}
{\cal P}_{\phi_1}(k) =  \frac{k^3}{2\pi^2} M_{\rm pl}^2 \, | \varphi_1({\bf k}) |^2  \simeq \left(\frac{H_I}{2\pi}\right)^2 .
\end{align}
These perturbations subsequently source isocurvature fluctuations during the radiation era, yielding
\begin{align}
{\cal P}_{\cal S}(k) =    \frac{k^3}{2\pi^2} |{\cal S}({\bf k})   |^2     =  \frac{4}{M_{\rm pl}^2 \varphi^2_{{\rm DM}}}  \cos^2(\theta_0 - \theta_1)  \, {\cal P}_{\phi_1} \big\vert_i \equiv A_{\cal S} \left( \frac{k}{k_*}\right)^{n_{\cal S} - 1}, 
\end{align}
where $k_*$ is the pivot scale. As a direct consequence of the scale invariance of ${\cal P}_{\phi_1}$, the resulting isocurvature spectrum is also nearly scale invariant, $n_{\cal S} \simeq 1$, with amplitude 
\begin{equation}
\label{eq:isocurvature-result}
 A_{\cal S} \equiv \frac{H_I^2}{\pi^2 M_{\rm pl}^2 \varphi_{{\rm DM}}^2 }  \cos^2(\theta_0 - \theta_1) \, .
\end{equation}

Precision measurements of the CMB by {\it Planck} place strong constraints on primordial isocurvature perturbations~\cite{Planck:2018jri}. In particular, {\it Planck} bounds the fractional contribution of isocurvature modes to the total primordial power at the pivot scale $k_* = 0.05$ Mpc$^{-1}$, finding $\beta_{\rm iso}(k_*) < 0.038$ for uncorrelated isocurvature with a nearly scale-invariant spectrum. 
Using the measured scalar amplitude $A_s = 2 \times 10^{-9}$, this implies the constraint $A_{\cal S} \lesssim 0.038 A_{s}$, which we use throughout in constraining the parameter space of the model.

%%%%%%%%%%%%%%%%%%%%%%%%%%%%%%%%%%%%%%%%
%%%%%%%%%%%%%%%%%%%%%%%%%%%%%%%%%%%%%%%%
\section{Results and Discussion}
\label{sec:results}
\begin{figure}
    \centering
    \includegraphics[width=0.9\linewidth]{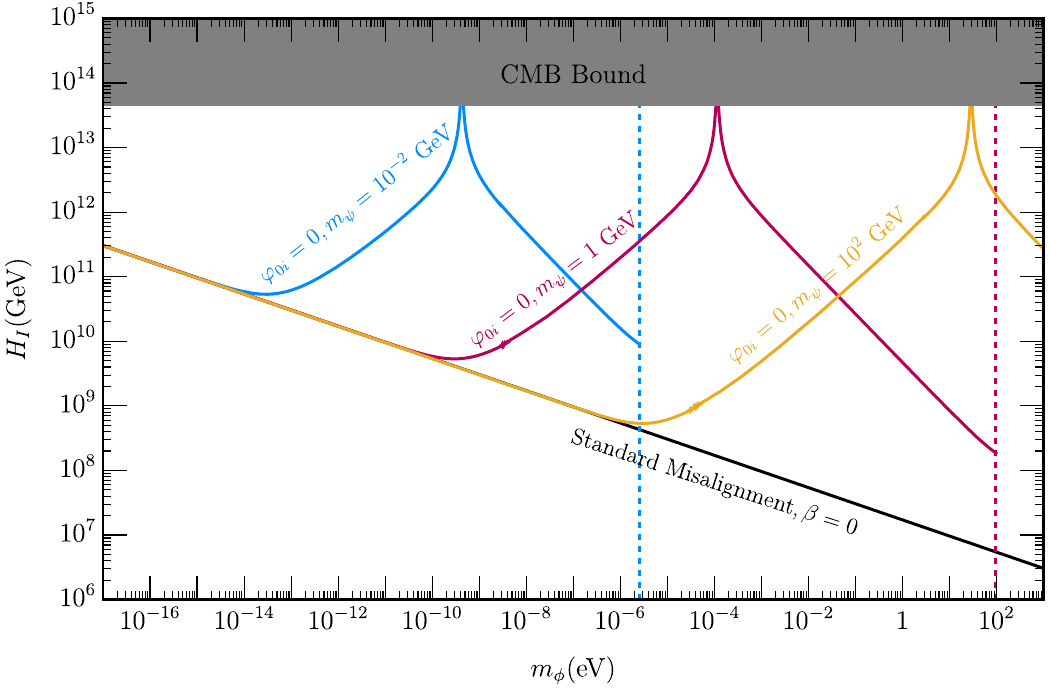}
    \caption{The upper bound on $H_{I}$ as a function of $m_\phi$ in thermal misalignment for several choices of fermion masses, $m_{\psi} = 10^{2}~\GeV$ (yellow), $1~\GeV$ (magenta) and  $10^{-2}~\GeV$ (blue), and for standard misalignment (black). For thermal misalignment, the upper bound is relaxed relative to standard misalignment over a broad scalar-mass range, with the precise interval depending on $m_\psi$. Vertical dashed lines indicate where the Born approximation fails (Eq.~(\ref{eq:Born-validity-f})). 
    The gray band indicates the Planck + BICEP  upper bound $H_{I} \leq 4.7 \times 10^{13} \GeV$ \cite{BICEP:2021xfz} from the non-observation of CMB B-modes associated with tensor fluctuations produced during inflation. 
    }
    \label{fig:HIplot}
\end{figure}

We now consider several limiting cases of the induced phase angles $\theta_0$ and $\theta_1$, Eqs.~(\ref{eq:varphi-0-osc-amp},\ref{eq:f-osc}), along with Eq.~(\ref{eq:isocurvature-result}) to obtain approximate expressions for the final isocurvature amplitude. 
First we will consider the standard misalignment ({\rm sm}) limit in which we set $\beta = 0$ and allow $\varphi_{0i} \neq 0$.
In this limit we get $\theta_{0}, \theta_{1} = 0$. We reproduce the well known isocurvature result (see, e.g., \cite{Tenkanen:2019aij}) \footnote{The same well-known result is obtained for QCD axion DM up to anharmonic corrections \cite{Seckel:1985tj,Linde:1987bx,Lyth:1989pb,Turner:1990uz,Lyth:1992tx,Fox:2004kb,Kobayashi:2013nva}}
\begin{align}
    \label{eq:iso-SM}
    A_{\cal S, {\rm sm}} = \frac{H_I^2}{\pi^2 M_{\rm pl}^2 \varphi_{{\rm DM}}^2 } \, .
\end{align}
Imposing the isocurvature constraint $A_{\cal S} \lesssim 0.038 A_{s}$ we obtain an upper bound on $H_{I}$ given by 
\begin{align}
\label{eq:Hbound}
H_{I} \leq 1.7 \times 10^7 \, {\rm GeV} \,  \left( \frac{1 \, {\rm eV}}{m_\phi}\right)^{1/4} \left(\frac{g_{*,{\rm osc}}}{106.75}\right)^{1/8}.
\end{align}
The resulting limit is shown in  Fig.~\ref{fig:HIplot}.

Next we consider the thermal misalignment ({\rm tm}) limit in which $\varphi_{0i} = 0$ and $\beta \neq 0$. 
The induced phase angles associated with the zero mode and superhorizon perturbations can be written as 
\begin{align}
    \label{eq:thetatm}
    \theta_0 &=  -\tan^{-1}\left(\frac{B(x_{\psi})}{A(x_\psi)}\right) = 
    \tan^{-1} \left(
    \frac{  F_J(x_\psi)
    +\frac{g_\psi}{2\pi}\frac{\beta^2}{\kappa^2}\Big[F_{JYJ}(x_\psi)- F_{JJY}(x_\psi)\Big]}{ 
    F_Y(x_\psi)  +\frac{g_\psi}{2\pi}\frac{\beta^2}{\kappa^2}\Big[ F_{YYJ}(x_\psi)-F_{JYY}(x_\psi)\Big]}
    \right),\\
    \theta_1 &= - \tan^{-1}\left(\frac{ D(x_\psi)}{ C(x_\psi)}\right) = -\tan^{-1} \left(\frac{\frac{g_\psi}{2\pi}\frac{\beta^2}{\kappa^2}\,{F}_{JJ}(x_\psi)}{1
- \frac{g_\psi}{2\pi}\frac{\beta^2}{\kappa^2}\,{F}_{JY}(x_\psi)}\right) \simeq - \frac{g_{\psi}}{2\pi} \frac{\beta^{2}}{\kappa^{2}} {F}_{JJ}(x_\psi).
\end{align}
In the last expression for $\theta_{1}$ we have restricted to the domain where the Born approximation, Eq.~(\ref{eq:Born-validity-f}), is valid. We also note that  $\theta_{1} \ll 1$ throughout this regime, i.e., the phase of the superhorizon perturbations is similar in standard and thermal misalignment.

To further understand the behavior of isocurvature perturbations in thermal misalignment, we examine the two regimes $\kappa \ll 1$ ($x_{\psi} \ll 1$) 
and $\kappa \gg 1$ ($x_{\psi} \gg 1$). 
Taking first the small $\kappa$ (small $x_\psi$) limit, Region 2, we make use of the asymptotic forms of the integrals given in Table~\ref{tab:integrals} and fix the coupling $\beta$ to match the observed DM relic  using Eqs.~(\ref{eq-phi-osc-tm},\ref{eq:phi-i-DM}), obtaining
\begin{align}
    \label{eq:thetatmreg1}
    \theta_0 & \simeq \frac{F_J(x_\psi)}{F_Y(x_\psi)} \simeq -1.63 x_\psi^{1/2}, \\
     \theta_1 & \simeq - 10^{-9}\left(\frac{100\, \rm GeV}{m_\psi}\right) \left(\frac{g_{*,{\rm osc}}}{106.75}\right)
      \frac{F_{JJ}(x_\psi)}{x_\psi^{1/2}[F_Y(x_\psi)]^2} \simeq - 10^{-9} \left(\frac{100\, \rm GeV}{m_\psi}\right) \left(\frac{g_{*,{\rm osc}}}{106.75}\right).
\end{align}
Both of the induced phase angles $\theta_0$ and $\theta_1$ are parametrically small, implying that $\cos(\theta_0-\theta_1) \simeq 1$. Therefore, for small $\kappa$ the isocurvature amplitude in thermal misalignment coincides with the standard misalignment result, Eq.~(\ref{eq:iso-SM}). This can also be understood from Fig.~\ref{fig:phievol} which shows that the phase of the late-time zero mode oscillations in thermal misalignment is nearly identical to that in standard misalignment.

The $\kappa \gg 1$ $(x_\psi \gg 1)$ limit, Region 1, exhibits more interesting behavior. Again making use of the asymptotic forms of the integrals presented in Table~\ref{tab:integrals} and setting the coupling $\beta$ to reproduce the correct relic abundance using Eqs.~(\ref{eq-phi-osc-tm},\ref{eq:phi-i-DM}), we obtain
\begin{align}
    \label{eq:thetazerotmreg2}
    \theta_{0} &= -\tan^{-1}\left(\frac{B(x_{\psi})}{A(x_{\psi})}\right) \simeq -\frac{\pi}{2} + \frac{A(x_{\psi})}{B(x_{\psi})}  \\
    &\simeq -\frac{\pi}{2}-\frac{{F}_Y(x_\psi) +\frac{g_\psi}{2\pi}\frac{\beta^2}{\kappa^2}\Big[{F}_{YYJ}(x_\psi)-{F}_{JYY}(x_\psi) \Big]}{{F}_J(x_\psi)} \nonumber \\
    & \simeq -\frac{\pi}{2} + \frac{(0.234+0.127 \log x_\psi)}{x_\psi}+
      10^{-9} \left(\frac{100\, \rm GeV}{m_\psi}\right) \left(\frac{g_{*,{\rm osc}}}{106.75}\right) (-1.61+1.15 \log x_\psi) x_\psi^{1/2}, \nonumber \\
    \label{eq:thetaonetmreg2}
       \theta_1 & \simeq - 10^{-9}\left(\frac{100\, \rm GeV}{m_\psi}\right) \left(\frac{g_{*,{\rm osc}}}{106.75}\right) 
      \frac{F_{JJ}(x_\psi)}{x_\psi^{1/2}[F_J(x_\psi)]^2}  \\
      & \simeq - 10^{-9} \left(\frac{100\, \rm GeV}{m_\psi}\right) \left(\frac{g_{*,{\rm osc}}}{106.75}\right)(0.296+1.15 \log x_\psi)x_\psi^{1/2} . \nonumber
\end{align}
For $\kappa\gg 1$ ($x_\psi \gg 1$), the thermal potential is unsuppressed at the onset of oscillations ($3H = m_{\phi}, x_{\rm osc} = 3/2$). 
As a result, the zero mode is generically not released from rest as in standard misalignment, but rather has a substantial velocity at $x_{\rm osc}$.
This results in a late-time oscillatory solution with a phase that is shifted by approximately $-\pi/2$ relative to the standard misalignment solution, which is captured by the leading term $\theta_0 \simeq - \pi/2$ in Eq.~(\ref{eq:thetazerotmreg2}).
The remaining terms in $\theta_0$ and $\theta_1$ are subleading corrections, controlled by the large-$x_\psi$ asymptotics and by restricting to the parameter space where the Born approximation is valid. In particular, imposing the Born-validity condition in Eq.~(\ref{eq:Born-validity-f}) ensures these corrections remain small, i.e., $|\theta_0 + \pi/2| \ll 1$ and $|\theta_1| \ll 1$. 

\begin{figure}
    \centering
    \includegraphics[width=0.9\linewidth]{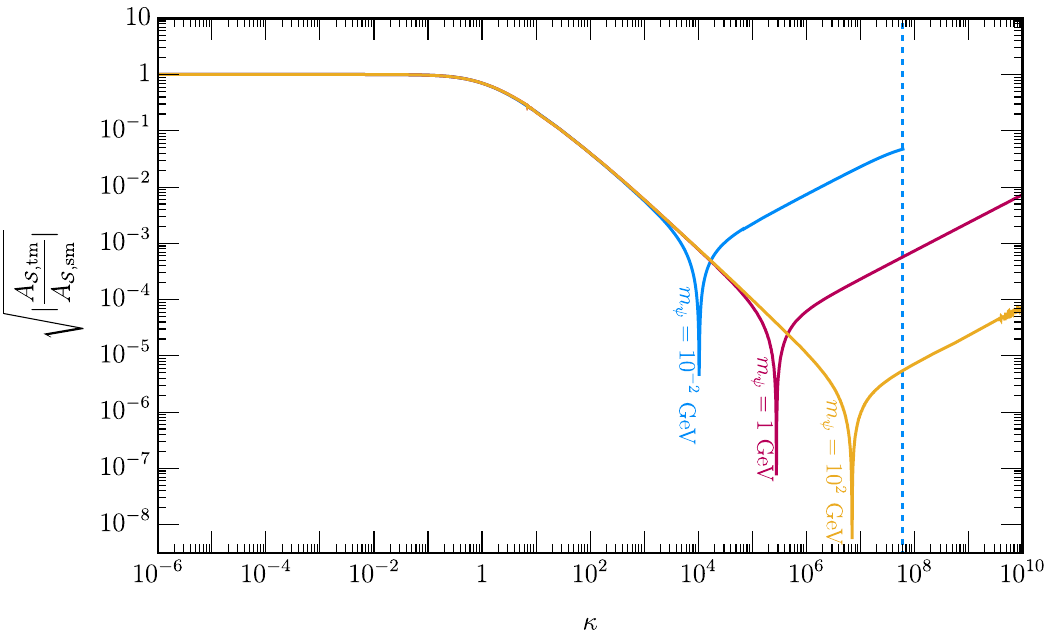}
    \caption{Square root of the isocurvature amplitude ratio between thermal misalignment and standard misalignment, $ \sqrt{|A_{\cal S, {\rm tm}}/{ A_{\cal S, {\rm sm}}}|}$ obtained using using Eqs.~(\ref{eq:isocurvature-result},    \ref{eq:iso-SM}). We have chosen three representative fermion masses, $m_{\psi} = 10^{2}~\GeV$ (yellow), $1~\GeV$ (magenta), $10^{-2}~\GeV$ (blue). For $m_{\psi} = 10^{-2} \,\GeV$ the vertical dashed line indicates the values of $\kappa$ above which the Born approximation fails (see Eq.~(\ref{eq:Born-validity-f})). The isocurvature in thermal misalignment exhibits the expected cosine dependence, with the amplitude ratio decreasing to zero before rising again. We can see the cosine behavior for isocurvature in thermal misalignment, as magnitude isocurvature ratio decreases to be exactly zero and increases again.
    }
    \label{fig:ratioplot}
\end{figure}

Using these results, the isocurvature amplitude for $\kappa \gg 1$ may be written as
\begin{align}
    \label{eq:iso-tm-largexpsi}
    A_{\cal S, {\rm tm}} & \simeq A_{\cal S, {\rm sm}}  \times \left[-\frac{{F}_Y(x_\psi)}{{F}_J(x_\psi)} -\frac{g_\psi}{2\pi}\frac{\beta^2}{\kappa^2}\left(\frac{\Big[{F}_{YYJ}(x_\psi)-{F}_{JYY}(x_\psi) \Big]}{{F}_J(x_\psi)} -{F}_{JJ}(x_\psi)\right)\right]^{2} \\
    & = A_{\cal S, {\rm sm}}  \times \left[  \frac{(0.234+0.127 \log x_\psi)}{x_\psi}
    -1.90 \times  10^{-9} \left(\frac{100\, \rm GeV}{m_\psi}\right) \left(\frac{g_{*,{\rm osc}}}{106.75}\right) x_\psi^{1/2}  \right]^2 .\nonumber
\end{align}
The factor in brackets in Eq.~(\ref{eq:iso-tm-largexpsi}) controls the ratio of isocurvature amplitudes, $\sqrt{| A_{\cal S, {\rm tm}}/ A_{\cal S, {\rm sm}}|} \simeq |\cos(\theta_{0}-\theta_{1})|$. For intermediate $x_\psi \gg 1$, the first term $\sim (c_1 +c_2 \log x_\psi)/x_\psi$  dominates, so the ratio  $\sqrt{| A_{\cal S, {\rm tm}}/ A_{\cal S, {\rm sm}}|}$  decreases with increasing $x_\psi$ and the isocurvature amplitude is increasingly suppressed. At larger $x_\psi$, the second term grows as $x_\psi^{1/2}$, leading to a value $x_\psi  = \hat x_\psi$ where the bracketed term vanishes and hence $A_{\cal S, {\rm tm}} \rightarrow 0$, given approximately by 
\begin{align}
    \label{eq:xpsi-zero}
 \hat  x_\psi & \simeq 1.26 \times 10^5  \left(\frac{m_\psi}{100\, \rm GeV}\right)^{2/3} \left(\frac{106.75}{g_{*,{\rm osc}}}\right)^{2/3} 
\bigg\{  20.4 + \log\left[   \left(\frac{m_\psi}{100\, \rm GeV}\right) \left(\frac{106.75}{g_{*,{\rm osc}}}\right) \right] \nonumber   \\
&+ \log\left[  20.4 + \log\left[   \left(\frac{m_\psi}{100\, \rm GeV}\right) \left(\frac{106.75}{g_{*,{\rm osc}}}\right) \right]   \right]
      \bigg\}^{2/3}.
\end{align}
Physically, this corresponds to a relative induced phase between the zero mode and its perturbation that yields complete cancellation of the isocurvature amplitude. 
 For even larger $x_\psi$, the $x_\psi^{1/2}$ 
term dominates and ${\cal S_{\rm ratio}}$ increases until the Born-validity bound in Eq.~(\ref{eq:Born-validity-f}) is reached. This behavior, including the single zero crossing, is shown in Fig.~\ref{fig:ratioplot}. We have numerically verified our analytical results for several benchmarks.

The suppression of isocurvature relative to standard misalignment permits larger values of $H_I$ without running afoul of observational isocurvature bounds. This enlarges the viable parameter space, making light scalar DM compatible with high-scale inflation. 
Since the isocurvature bound is appreciably relaxed only for  $\kappa \equiv m_{\phi} M_{\rm pl}/m^{2}_{\psi} \gg 1$,  the impacted range of scalar masses depends on the fermion mass $m_\psi$. Fig.~\ref{fig:HIplot} illustrates the resulting enlargement of viable parameter space for several representative choices of fermion masses.

The above result was derived in the limit $\varphi_{0i} =0$. However, for a small but finite $\varphi_{0i} \ll \varphi_{{\rm DM}}$, 
the suppression factor in the brackets of  Eq.~(\ref{eq:iso-tm-largexpsi}) receives an additional contribution of $\varphi_{0i}/\varphi_{{\rm DM}}$,
setting a floor on the extent to which isocurvature can be suppressed.

%%%%%%%%%%%%%%%%%%%%%%%%%%%%%%%%%%%%%%%%
%%%%%%%%%%%%%%%%%%%%%%%%%%%%%%%%%%%%%%%%
\section{Outlook}
\label{sec:conclusions}

In this work we have studied the generation of DM isocurvature perturbations in the framework of thermal misalignment. We showed that, unlike in standard misalignment, the isocurvature signal can be naturally suppressed in regions of parameter space where the scalar begins oscillating while the thermal potential still dominates. Physically, this suppression arises because the thermal potential gives the zero mode a nonzero initial velocity at the onset of oscillations, producing a late-time phase offset relative to the superhorizon perturbations and thereby reducing the final DM density contrast. Thermal misalignment therefore opens new parameter space for light scalar DM consistent with CMB isocurvature bounds and with well-motivated models of high-scale inflation.

There are several directions that deserve further study.  The thermal potential induces energy and momentum exchange between $\phi$ and the thermal bath, modifying the evolution of both adiabatic and isocurvature perturbations. In this language, the isocurvature suppression found here can be understood as a consequence of energy and momentum transfer between the two sectors~\cite{Malik:2002jb,Gupta:2003jc,Malik:2004tf,Bellomo:2022qbx,racco2023freeze,Strumia:2022qvj}, a perspective that we plan to develop further in future work~\cite{Batell:2026xxx}. 
It would also be interesting to study the impact of thermal misalignment on the matter power spectrum at large $k$. 
One further direction is to explore other mechanisms beyond thermal misalignment which realize the phase off-set between the zero mode and superhorizon perturbations as a means to suppress DM isocurvature.

Finally, while we have focused here on a minimal realization involving a scalar interacting with a bath fermion via a Yukawa coupling, it would be valuable to extend the analysis to a broader class of models. Particularly interesting are scenarios in which scalar DM has shift-symmetry-breaking couplings to SM particles, potentially leading to additional experimental signatures that could test the relic-density prediction of thermal misalignment.

%%%%%%%%%%%%%%%%%%%%%%%%%%%%%%%%%%%%%%%%
%%%%%%%%%%%%%%%%%%%%%%%%%%%%%%%%%%%%%%%%
\acknowledgments

We would like to thank Priyesh Chakraborty, Chandrika Chandrashekhar, Wayne Hu, Soubhik Kumar, Rashmish Mishra, Sonia Paban, Matthew Reece and Sarunas Verner for helpful conversations. 
The work of BB is supported by the U.S. Department of Energy under Grant No. DE–SC0007914. 
The work of AG is supported by the GRASP initiative at Harvard University. SG acknowledges support from the National Science Foundation (NSF) under Grant No. PHY-2413016. 
The work of MR is supported in part by the U.S. Department of Energy under Grant No. DE-SC0010813.

%%%%%%%%%%%%%%%%%%%%%%%%%%%%%%%%%%%%%%%%
%%%%%%%%%%%%%%%%%%%%%%%%%%%%%%%%%%%%%%%%
\appendix

%%%%%%%%%%%%%%%%%%%%%%%%%%%%%%%%%%%%%%%%
%%%%%%%%%%%%%%%%%%%%%%%%%%%%%%%%%%%%%%%%

% \section{Appendix ...}
%\bibliographystyle{apsrev4-1}
\bibliographystyle{jhep}
\bibliography{refs}
%%%%%%%%%%%%%%%%%%%%%%%%
\end{document}